\documentclass[sigconf,screen]{acmart}

\makeatletter
\def\subsubsection{\@startsection{subsubsection}{3}%
  \z@{.5\linespacing\@plus.7\linespacing}{.1\linespacing}%
  {\it\sf\bf}}
\makeatother

\makeatletter
\renewcommand\@formatdoi[1]{\ignorespaces}
\makeatother
\renewcommand\footnotetextcopyrightpermission[1]{}

\renewcommand\footnotetextcopyrightpermission[1]{} 
\usepackage{tcolorbox}
\usepackage{balance} 
\usepackage{booktabs} 
\usepackage{amsmath}
\usepackage{graphicx}
\usepackage[linesnumbered,ruled,vlined]{algorithm2e}
\usepackage{epstopdf}
\usepackage{url}
\usepackage{enumerate}
\usepackage[labelfont=bf]{caption}
\usepackage{xcolor}
\usepackage{color, colortbl}

\usepackage{pifont}

\newcommand{\sys}{DSM-DB}

\newcommand{\sect}{Sec.}

\usepackage{indentfirst}

\begin{document}
\pagestyle{plain} 

\title{The Case for Distributed Shared-Memory Databases \\with RDMA-Enabled Memory Disaggregation [Vision Paper]}

\author{Ruihong Wang \;\; Jianguo Wang \;\; $^\dag$Stratos Idreos \;\; $^\S$M. Tamer {\"{O}}zsu \;\; Walid G. Aref}
\affiliation{
\vspace{0.1cm}
 \institution{\textsf{Purdue University} \hspace{1cm} $^\dag$\textsf{Harvard University} \hspace{1cm} $^\S$\textsf{University of Waterloo}}
 \vspace{0.1cm}
  \city{\textit{\{wang4996; csjgwang; aref\}@purdue.edu} \hspace{0.5cm} $^\dag$\textit{stratos@seas.harvard.edu} \hspace{0.5cm} $^\dag$\textit{tamer.ozsu@uwaterloo.ca} }
  \country{}
}

\begin{abstract}
Memory disaggregation (MD) allows for scalable and elastic data center design  
by separating compute (CPU) from memory. With MD, compute and memory are no longer coupled into the same server box. Instead, they are connected to each other via ultra-fast networking such as RDMA. MD can bring many advantages, e.g., higher memory utilization, better independent scaling (of compute and memory), and lower cost of ownership. 

This paper makes the case that MD can fuel the next wave of innovation on 
database systems. 
We observe that MD revives the great debate of "shared what" in 
the database community. 
We envision that \textit{distributed shared-memory databases (\sys{}, for short)} -- that have not received much attention before -- can be promising in the future with MD. We present a list of challenges and opportunities that can inspire next steps in system design making the case for \sys{}.
%
%
\end{abstract}

\maketitle

\settopmatter{printfolios=true}

\section{Introduction}\label{sec:intro}

\begin{figure}[tbp]
\centering
\includegraphics[width=0.35\textwidth]{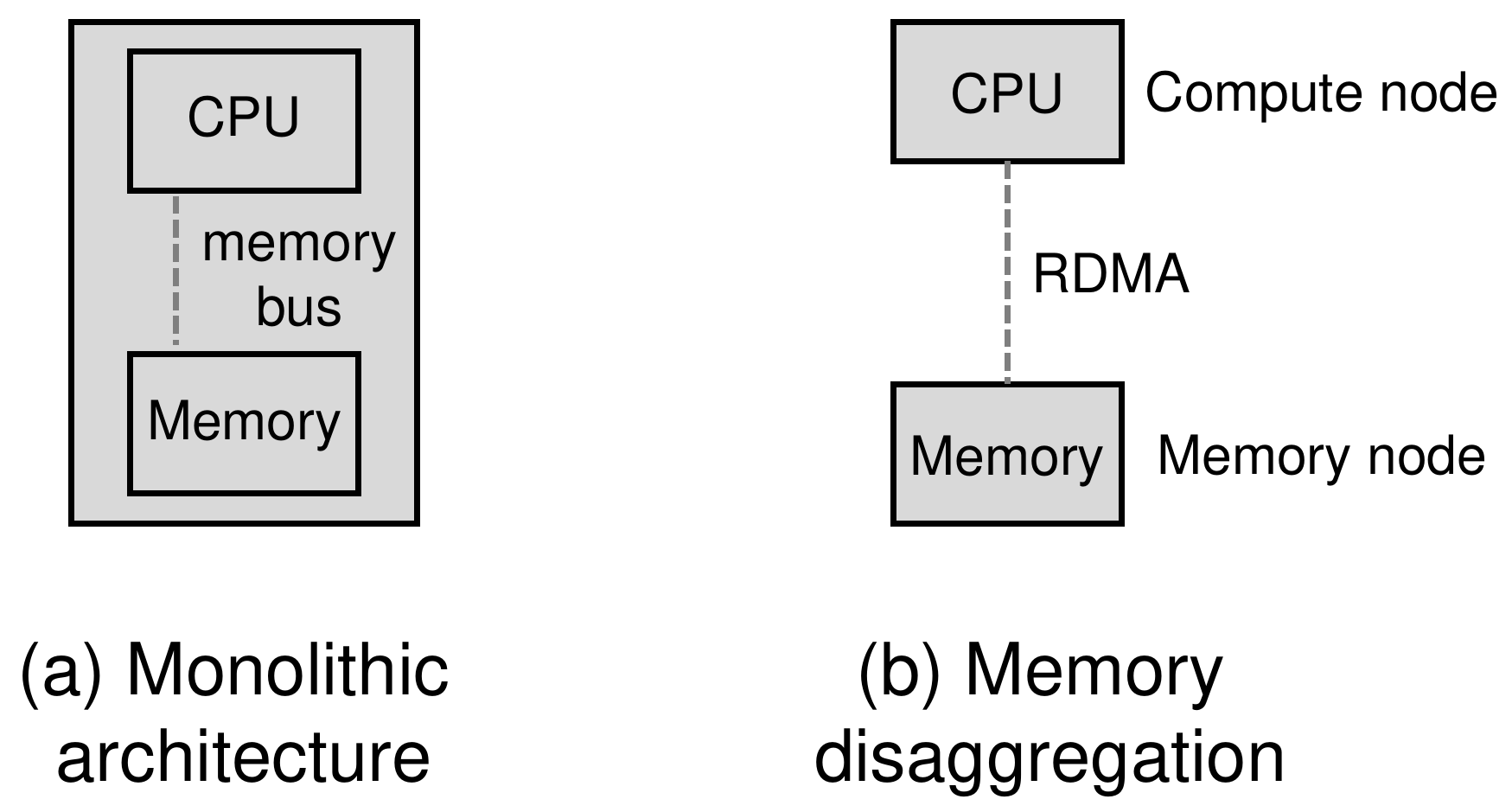}
\caption{Monolithic architecture vs. memory disaggregation}\label{fig:disaggregated}
\end{figure}

\underline{M}emory \underline{D}isaggregation, MD for short,
is emerging
as a promising architecture in modern data centers, especially in the cloud~\cite{PolarDBServerless21,ZhangCIDR20,Disaggregation21,ZhangVLDB20}. MD enables data center design based on independent pools of \emph{compute nodes} and \emph{memory nodes} that are physically separated but are connected via ultra-fast RDMA networks (Figure~\ref{fig:disaggregated}b)~\cite{PolarDBServerless21,ZhangCIDR20,Disaggregation21,ZhangVLDB20}. A critical enabler for MD is hardware advances, especially networking technology. RDMA, e.g., Mellanox Connectx-6~\cite{Mellanox}, 
achieves 0.8 $\mu$sec latency and 200Gb/s throughput -- close to local memory performance though there is still a gap. 

Overall, this is in contrast to traditional data centers that consist of a collection of monolithic "converged" servers, where compute and memory are tightly coupled in the same physical servers (Figure~\ref{fig:disaggregated}a). With MD, compute nodes focus on computation while  memory nodes are dedicated for provisioning memory.\footnote{With storage disaggregation, there are also dedicated storage nodes, but  this paper focuses on memory disaggregation.} The compute nodes may include small amounts of memory and memory nodes may have some compute capability to run simple control software. Thus, a compute node has strong computing power (e.g., 100s of CPU cores) but limited local memory (e.g., a few GBs) while a memory node has weak computing capability (e.g., a few CPU cores) but abundant memory (e.g., 100s of GBs) to store data~\cite{ZhangCIDR20,ZhangVLDB20}.

MD brings many advantages for data centers~\cite{ZhangCIDR20,ZhangVLDB20,PolarDBServerless21,Disaggregation21,IBMCloud}. (1)~MD results in higher memory utilization with less  fragmentation due to memory pooling. This translates into lower memory consumption and lower total cost of ownership (TCO) as memory is still an expensive resource. 
(2)~MD provides independent elastic scaling of compute and memory, which is very useful in the cloud. It allows users to request instances with arbitrary combinations of compute and memory that existing monolithic servers fail to provide. Also, compute or memory can be elastically adjusted with workload changes. (3)~MD achieves better reliability and lower operational cost because compute and memory failures and upgrades are independent, and do not affect each other. (4)~MD provides users virtually a near-infinite pool of memory for applications.

For all the above reasons, MD is receiving increasingly more attention from industry. For example, PolarDB, a cloud-native DBMS, relies on MD to improve memory utilization, and supports independent scaling~\cite{PolarDBServerless21,Disaggregation21}. Microsoft Azure has started to build an MD system in the cloud to improve memory utilization~\cite{AzureMD2022}. IBM Cloud provides MD in cloud data centers to significantly reduce cost, and achieve better reliability~\cite{IBMCloud}. Intel RSD (Rack Scale Design)~\cite{IntelRSD} and HP "The Machine"~\cite{HPTheMachine} also support MD at rack scale. 

\textbf{Position}. We argue that the next wave in database system innovation should be shared-memory designs enabled by RDMA-based MD. Similar to previous design evolution in decoupling system components 
in favor of scalability and elasticity, we believe that MD is the key to move database design to the next frontier. 
We present the challenges and opportunities in realizing distributed shared-memory databases (\sys{}) with MD. This vision paper focuses on OLTP main-memory databases for high performance.


\section{Vision and Contributions}\label{sec:arch}

\begin{sloppypar}
\textbf{Novelty and Architectural Evolution}. 
\sys{} architectures have been proposed in the 1980s~\cite{StonebrakerSharedNothing86}. However, slow networking at the time 
has made these architectures infeasible due to the slow access of remote memory. Thus, existing distributed DBMSs are mostly shared-nothing, or recently, shared-storage architectures. For decades, the shared-nothing architecture has been regarded as the "gold standard" in distributed DBMSs due to their high performance especially in supporting single-shard queries~\cite{StonebrakerSharedNothing86,StonebrakerSharedNothing2011}. Examples include MySQL Cluster, PostgreSQL Citus~\cite{Citus21}, Teradata, MemSQL~\cite{MemSQL16}, VoltDB~\cite{VoltDB}, SQL Server PDW, and Greenplum. 
As the cloud becomes prevalent, shared-storage DBMSs start to gain attention because they can best leverage the cloud infrastructure of storage disaggregation. Examples include Amazon Aurora~\cite{Aurora17}, Google AlloyDB~\cite{AlloyDB}, Alibaba PolarDB~\cite{PolarFS18}, Alibaba AnalyticDB~\cite{AnalyticDB19}, Microsoft Socrates~\cite{Socrates19}, Microsoft Polaris~\cite{POLARIS20}, Huawei Taurus~\cite{Taurus20}, and  Snowflake~\cite{Snowflake16}. The shared-storage architecture is compelling in the cloud because it can support independent scaling of compute and storage, better elasticity, and fast crash recovery -- all are important design considerations for the cloud. However, the shared-storage architecture still suffers from high memory consumption and high cost because compute and memory remain tightly coupled.
This vision paper is a natural step forward arguing for a \sys{} architecture driven by MD and ultra-fast RDMA networking. 
\end{sloppypar}

\begin{figure}[tbp]
\centering
\includegraphics[width=0.35\textwidth]{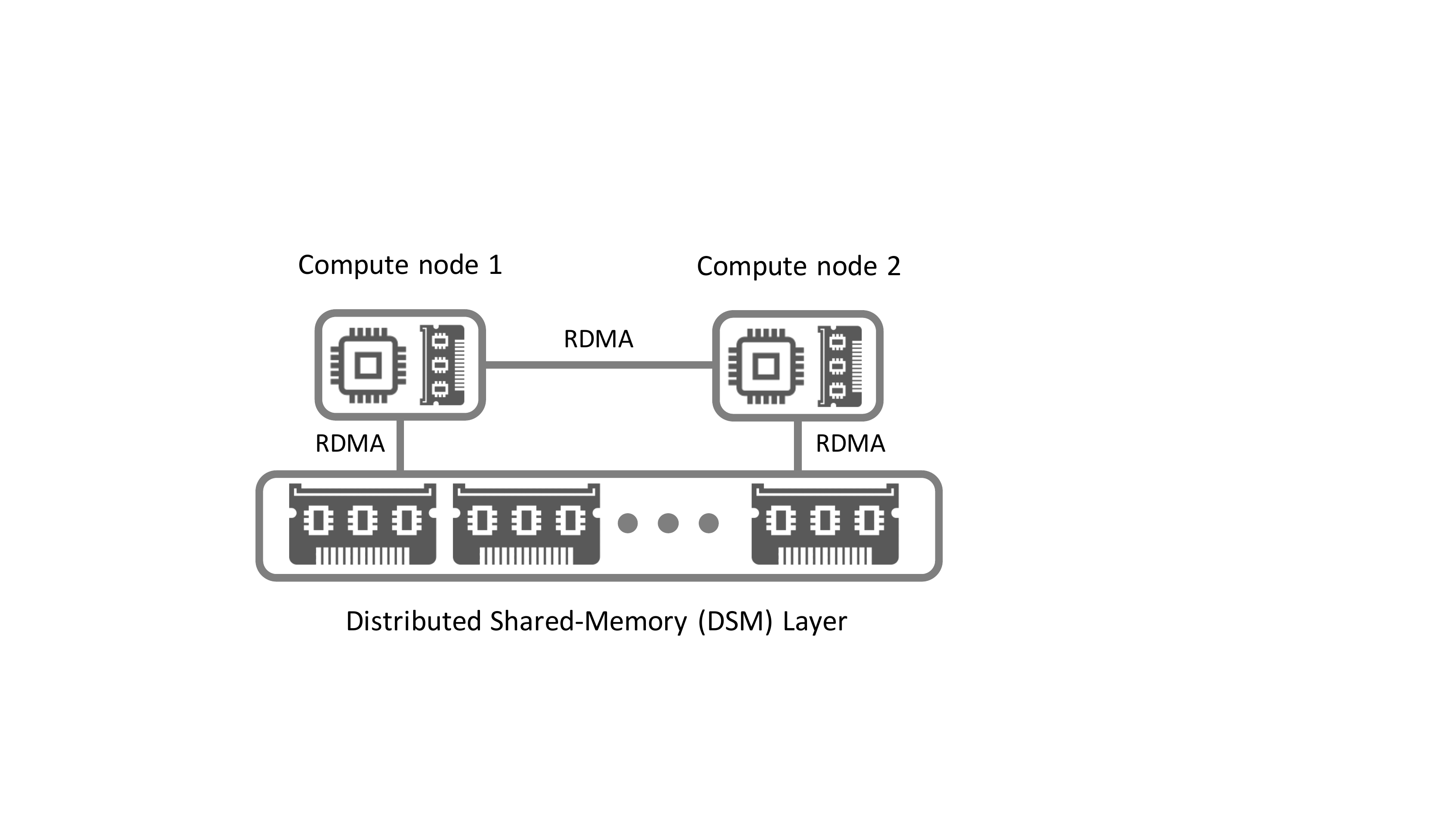}
\caption{\sys{} with memory disaggregation (MD)}\label{fig:arch}
\end{figure}

\textbf{Distributed Shared-Memory Database Architecture (\sys{}) }. Figure~\ref{fig:arch} shows the \sys{} architecture  that separates compute and memory nodes. Memory nodes form a distributed shared-memory  (DSM) layer that is  shared by compute nodes via an ultra-fast RDMA network. Compute nodes have high computing power with limited local memory while memory nodes have large memory capacity with limited computing power. 
Compute nodes can communicate with each other via RDMA. 
\sys{} is an OLTP distributed main-memory DBMS that stores data in the DSM layer with hot data being cached in the compute nodes' local memories.

\textbf{Expected Benefits}. \sys{} has many advantages over distributed shared-storage or shared-nothing databases. 
(1)~\sys{} has independent elasticity of compute and memory due to MD. 
(2)~\sys{} incurs lower total cost of ownership by increasing memory utilization as the majority of data is stored in the distributed shared-memory while each compute node only keeps small amounts of local private memory. 
(3)~\sys{} supports independent failure of compute and memory nodes to achieve high availability.
(4)~\sys{} is more robust to query and data skew that is difficult to handle in distributed shared-nothing databases as data can be easily resharded in DSM. 
(5)~\sys{} can achieve better scalability with  multi-masters
(\sect{}~\ref{sec:cc}) that existing distributed shared-storage databases would not have.
Note that the performance issues in \sys{} due to MD can be mitigated (see \sect{}~\ref{sec:drawbacks}).

\textbf{Contributions}. 
This paper presents a list of challenges in realizing \sys{}. 
(1)~\textit{DSM Layer}  (\sect{}~\ref{sec:dsm}): This layer needs to provide high durability and availability at a low cost. Also, it needs to expose a rich API to the DBMS. 
(2)~\textit{Concurrency Control} (\sect{}~\ref{sec:cc}): Multi-node concurrency control is challenging due to the lack of cache-coherence across compute nodes.
(3) \textit{Buffer Management} (\sect{}~\ref{sec:buf}): Managing data movement between local and remote memory requires rethinking how to leverage fast RDMA networking; otherwise, due to the relatively high speeds of these layers in contrast to disk, the software layer can become the new bottleneck. 
(4) \textit{Index Design} (\sect{}~\ref{sec:index}): It is non-trivial to develop indexes that  fully exploit RDMA, and that support highly-concurrent accesses. 

\section{Distributed Shared-Memory}\label{sec:dsm}

\begin{sloppypar}
The goal for having  distributed shared-memory (DSM) in \sys{} is to  manage a cluster of memory nodes (each  provisioning large memory) and provide unified memory space with the necessary APIs for DBMSs to build on. Compute nodes access DSM  via these APIs.
\sys{} introduces a DSM layer to encapsulate all  memory management details and hide them from compute nodes for two reasons. 
First, it enables independent elasticity of compute vs. memory. Second, it simplifies system design because each compute node sees a unified infinite memory space and only focuses on query processing logic inside a compute node without worrying about the complicated memory management underneath.
\end{sloppypar}

\textbf{Challenge \#1: Exposing Abstract APIs}. What APIs best support DBMSs? The APIs must include not only basic memory access APIs, e.g., memory allocation,  deallocation,  read, and write, but also database functions for near-data computing. 
Thus, DSM will provide the following APIs to support database applications. 

\textit{Memory Allocation APIs}. DSM will provide memory allocation APIs similar to those in a single memory node, e.g.,  memory allocation, deallocation, and reallocation. However, \textbf{memory address representation} is challenging as it cannot be the physical address allocated via a programming language, e.g., \texttt{malloc} in C++. 
If a memory node crashes then recovers, the memory space changes and the old address cannot refer to the new memory. Thus, the memory address must be a logical address, e.g., 
virtual node ID and offset. 
To allocate memory efficiently and reduce memory fragmentation, \sys{} can  allocate a giant continuous memory space and keep track of memory usage in user space~\cite{TaranovGH21}.

\textit{Data Transmission APIs}. 
These APIs provide one- and two-sided RDMA, memory access (read/write), and atomic operations for concurrency.

\textit{Function Offloading APIs}. These APIs will push down certain database functions to memory nodes that leverage the computing resources in DSM to reduce data transfer.

\textbf{Challenge \#2:  Durability}. A single memory node is volatile.  DSM must be durable to ensure that committed data is not lost. At a minimum, a crash in a single memory node must not cause data loss.
Upon transaction commit, logs must be written to persistent storage. However, log persistence needs to be highly-performant at low cost so that it is not the bottleneck in main-memory databases. The following are possible directions to be explored.

\textit{Approach \#1}. One possibility is to write logs to durable storage as in main-memory databases~\cite{MMDB17}. \sys{} can choose cloud storage, e.g., AWS EBS and S3 are highly reliable with low cost, and can achieve 99.999\% and 99.999999999\% durability, respectively~\cite{AWSEBS, AWSS3}. Cloud storage can be viewed as  distributed shared storage that is accessible by all compute and memory nodes. Crash recovery is similar to that in main-memory databases~\cite{MMDB17}. However,  writing to cloud storage is relatively slow and is on the critical path for  transaction commit. The same problem arises in main-memory databases~\cite{MMDB17}. Thus, similar optimizations need to be revised for \sys{}, e.g., group commit~\cite{DeWittKOSSW84,GarciaMolinaS92}, command logging~\cite{MalviyaWMS14},  logging only base data but not indexes~\cite{Hekaton13}. For instance,  command logging in \sys{} cannot rebuild the same states upon crash because with multi-master, the system may not be able to determine the global transaction order in advance.

\textit{Approach \#2}. Another possibility is to follow RAMCloud~\cite{ramcloud09} that uses memory replication to emulate durable storage. It writes a log synchronously to $k$ different memory nodes ($k=3$ in RAMCloud) and a log write is considered "persistent" if all $k$ memory nodes successfully write the log to their own main-memories. If one node crashes, a new node may be identified and is restored from the $(k-1)$ replicated nodes. Compared to Approach \#1,  log persistence is fast as it does not involve disk. But it may not guarantee 100\% durability as the probability of all $k$ memory nodes crashing is not zero, and this could lead to data loss. Remedies could be battery-backed memory~\cite{GarciaMolinaS92} or persistent memory~\cite{ArulrajPP16}. More research is needed to investigate the interaction with RDMA.

\textbf{Challenge \#3:  Availability}. Main-memory is volatile and \sys{} can become unusable upon crash. The goal is to achieve reasonable availability to minimize downtime while taking a low (monetary) cost. 
A simple solution is to replicate data in different memory nodes to support high availability. This consumes memory, and hence is expensive. Another solution is to use erasure code~\cite{SathiamoorthyAPDVCB13,Hydra22} but the recovery process is long if there is a crash. The third solution is to follow the RAMCloud approach~\cite{ramcloud09} that stores data pages in main-memory only once to reduce memory consumption. To improve availability, RAMCloud periodically checkpoints data pages from memory nodes to persistent store (this can be cloud storage in \sys{}). If a memory node  crashes, its content can be recovered by accessing the persistent store and possibly replaying some of the logs. More research is required to speedup crash recovery since accessing cloud-based persistent storage is relatively slow.

\textbf{Existing Research}. 
Early works on DSM, e.g., \cite{Kai89,ProticTM96,NitzbergL91},  do not target MD. In those works, all nodes are homogeneous and  compute and memory nodes are not differentiated. Also, they do not use RDMA and do not support durability and availability. 
Existing distributed memory systems, e.g., GAM~\cite{GAM18}, NAM~\cite{ZamanianBKH17,BinnigCGKZ16}, FarM~\cite{FaRM14}, Redy~\cite{Redy22}, and Memcached~\cite{Memcached} do not provide durability or availability, and do not provide database-specific functions.

 RAMCloud~\cite{ramcloud09} is  memory-based and offers durability and availability but cannot be used as the DSM layer in \sys{}. First, RAMCloud has no memory APIs (e.g., memory allocation and memory read/write). Instead, it has key-value APIs~\cite{OngaroRSOR11}. Second, RAMCloud does not provide database-specific functions, e.g., offloading. Third, RAMCloud assumes TCP/IP  rather than  RDMA networking.

\begin{figure*}[tbp]
\centering
\includegraphics[width=0.8\textwidth]{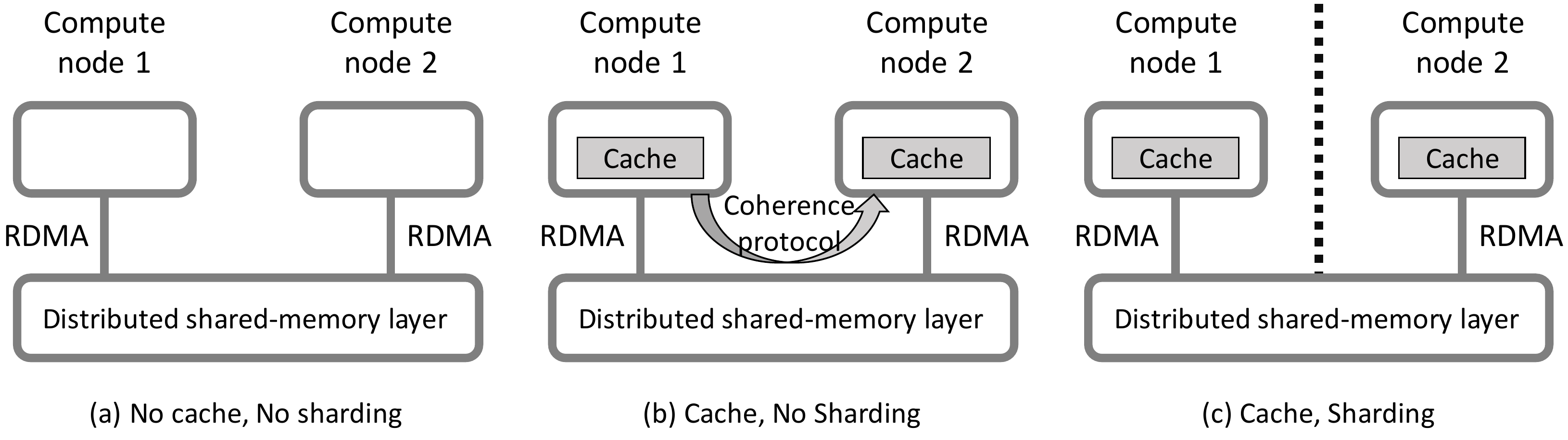}
\caption{Concurrency control design tradeoffs in \sys{}}\label{fig:concurrency}
\end{figure*}

\vspace{-10pt}
\section{Concurrency Control}\label{sec:cc}

Since the compute nodes share access to the memory nodes in \sys{}, their concurrent accesses to the memory pool need to be protected through a scalable concurrency control (CC) protocol. 
There are a few new challenges when compared to the CC protocols in the multi-core architecture, e.g.,  ~\cite{AbyssCC2014,MOCC16,BangMPB20,TanabeHKT20}.

\textbf{Challenge \#4: The Cache Coherence Challenge}. 
In \sys{}, there is no hardware-level cache coherence among the compute nodes. If a compute node updates a data item, another compute node may not see the update immediately. This is different from the conventional single-server multi-core architecture  (with different CPU cores sharing the memory) because hardware-level cache coherence among different CPU cores is natively provided.

Cache coherence is important as it affects the design decision on whether or not to use the local buffer memory in compute nodes. If local buffers are used, then cache coherence needs to be addressed at the software-level, which adds performance overhead. Otherwise, the cache coherence problem is bypassed at the expense of more remote accesses as the compute nodes will have to always access data from remote memory. 
A related design decision is whether or not to allow different compute nodes to access disjoint data partitions, i.e., sharding. With sharding, the cache coherence issue can be avoided when every data page is accessed by a single compute node. The following three approaches to address the cache coherence challenge need to be systematically evaluated.

\textit{Approach \#1: No Cache, No Sharding} (Figure~\ref{fig:concurrency}a). 
Recent RDMA-optimized CC techniques~\cite{ZamanianBKH17,RDMABtree19,wang2021rdma,Sherman2022}  follow this approach. A compute node reads and writes data remotely and does not store any data in local memory. Data is stored in DSM with a lock per data item.  Compute nodes use RDMA Compare \& Swap (CAS) to acquire a lock before accessing data. There is no cache coherence issue as no local data exists in a compute node. Because a compute node always accesses data remotely, this incurs performance overhead.

\textit{Approach \#2: Cache, No Sharding} (Figure~\ref{fig:concurrency}b).  Compute nodes leverage local memory to perform reads and writes. This may create cache coherence issues when two compute nodes update the same data locally. To resolve
conflicts, a software-level cache coherence protocol, e.g., \cite{GAM18,PolarDBServerless21}, is needed to broadcast changes made by a compute node. 
However, the effect of software overhead is unclear as many implementation details can affect performance, e.g., invalidation- vs. update-based, one- vs. two-sided RDMA, fetching missed data from neighboring compute nodes or from the shared-memory layer. To reduce  cache coherence overhead, a lazy cache coherence protocol can trade mutual consistency for performance.

\textit{Approach \#3: Cache, Sharding (Figure~\ref{fig:concurrency}c)}. 
%
In this approach, we perform \textit{logical sharding}, where each compute node maintains  sharding information (e.g., range information) of the data it is responsible for. 
CC is similar to the case in distributed shared-nothing databases, e.g., \cite{EvaDT17}. However, the difference is that a compute node does not store the entire data shard because of the limited local memory although it can cache some hot data. The advantage is that there is no cache coherence issue due to sharding. Also, this approach can best leverage local memory. Another advantage is that elasticity can be supported very well, e.g., if a new compute node is added, only the metadata (e.g., range information) is copied into the new node without physically moving data (due to logical sharding), and the obsolete data from the old compute nodes can be recycled asynchronously. The downside is in cross-shard transactions. However, this can be alleviated via dynamic resharding~\cite{LinC0OTW16,AbebeGD20} that is efficient in \sys{} since the DSM layer can transfer data quickly.

\textbf{Challenge \#5: Rethinking Distributed Commit}. 
Are the distributed commit protocols (e.g., 2PC)~\cite{OzsuV14} -- widely used in distributed shared-nothing databases -- still applicable in \sys{}? Observe that distributed commit may not always be relevant in \sys{}  depending on which architecture in Figure~\ref{fig:concurrency} is adopted. If \sys{} uses a no-sharding architecture (Figure~\ref{fig:concurrency}a or Figure~\ref{fig:concurrency}b), there is no need for distributed commit. If there is no sharding between different compute nodes, every compute node can read/write all the data. Thus,  every transaction will be executed by a single compute node, which does not require distributed commit.

In contrast, if \sys{} uses sharding   (Figure~\ref{fig:concurrency}c), distributed commit may become relevant. Each compute node is responsible for a shard of data pages and a transaction may access pages belonging to different compute nodes. More research is needed to leverage  RDMA primitives in distributed commit protocols. 
Notice that sharding data among memory nodes does not necessarily mandate  using 2PC. If a compute node uses one-sided RDMA to access memory nodes, it knows whether or not a write is successful. 

\textbf{Challenge \#6: Revisiting Existing CC Protocols}. 
In \sys{}, compute nodes access remote memory via RDMA. It is costly to implement locks over RDMA (compared with the conventional local locks). This has implications to both lock-based and non-lock-based CC protocols that need to be revisited in \sys{}.

\textit{Lock-based CC}.  
Lock-based concurrency control protocols, e.g., 2PL, rely on fine-grained locks to improve the level of concurrency, e.g., shared-exclusive locks, intention locks. 
It is challenging to implement these lock variants on RDMA and the implementation overhead varies. 
RDMA can only implement a simple exclusive spinlock \textit{within a single round trip} through the CAS atomic primitive. Advanced lock types require more RDMA round trips, e.g., an RDMA shared-exclusive lock needs at least 2 round trips.\footnote{One way to implement shared-exclusive locks is to use a lock metadata and an RDMA spinlock on the remote memory. The lock metadata records the number of holders/waiters on different lock modes. The RDMA spinlock guarantees atomicity when modifying the lock's metadata. It needs at least two RDMA round trips where Round \#1 is to acquire the lock and read the metadata and Round \#2 is to update the metadata and release the lock.}
Thus, it is unclear whether advanced lock types could be used for lock-based CC algorithms. It remains open if the allowed extra concurrency can offset the performance overhead of the advanced locks.

\textit{Non-lock-based CC}.  
RDMA can also impact non-lock-based concurrency control protocols, e.g., timestamp-based CC, MVCC, and optimistic CC though in a mild way. These protocols need latches using RDMA to exclusively access some shared states, e.g., global timestamps. Another related optimization is how to generate timestamps. One-sided RDMA (RDMA Fetch \& Add) is more preferable than two-sided RDMA in case that the centralized timestamp generator becomes a bottleneck. It is interesting to investigate other approaches (e.g., vector timestamp~\cite{ZamanianBKH17} and clock synchronization~\cite{wang2021rdma}). A systematic evaluation of different concurrency control protocols over RDMA is necessary.

\textbf{Challenge \#7: Supporting Massive Concurrency}. Prior work has examined concurrency control for 1000s of cores per compute node~\cite{AbyssCC2014,MOCC16,BangMPB20,TanabeHKT20}.  \sys{} can enable  ultra-high number of cores (e.g.,  millions) accessing the same shared-memory by supporting many compute nodes (e.g., 10s to 1000s). It would be interesting to re-evaluate and re-think CC protocols to support extreme concurrency, e.g., millions of cores. This may require distinguishing the local concurrency control (within the same compute node) and global concurrency control (across different compute nodes).

\textbf{Existing Research}. There are a number of existing works in concurrency control that are relevant to \sys{}.

\textit{Concurrency Control in Multi-Core Databases}. Most CC work, e.g., 2PL, MVCC, and OCC are designed for multi-core setups~\cite{AbyssCC2014, MOCC16, BangMPB20, TanabeHKT20}. They share the implicit assumption of having hardware-supported cache coherence, which does not exist in \sys{}.

\begin{sloppypar}

\textit{Concurrency Control in Distributed Shared-Storage Databases (DSS-DB)}. DSS-DBs, e.g.,  Aurora~\cite{Aurora17}, PolarDB~\cite{PolarFS18}, Socrates~\cite{Socrates19}, and Taurus~\cite{Taurus20} do not support concurrent transactions among multiple compute nodes in order to avoid conflicts. Instead, only the primary node can support writes (aka single-writer) while all the other nodes are replicas for read-only transactions. However, \sys{} can support multi-writers where every compute node supports writes to improve write scalability.
\end{sloppypar}

\begin{sloppypar}

\textit{Concurrency Control in Distributed Shared-Nothing Databases (DSN-DB)}. Concurrency control is extensively studied in DSN-DBs~\cite{EvaDT17, Calvin12, Granola12}. While it is possible to adapt these techniques into \sys{} by sharding the data among different compute nodes, there are two differences in \sys{} that may inspire innovation. 
(1)~A compute node does not physically store the entire data shard due to the limited local memory. 
(2)~Data/state can be moved quickly among compute nodes via the DSM layer.
\end{sloppypar}

\textit{Other Works}. PolarDB considers MD~\cite{Disaggregation21,PolarDBServerless21} but 
only the primary node can write data. There are concurrency control protocols optimized for RDMA, e.g., ~\cite{ZamanianBKH17,RDMABtree19,wang2021rdma,Sherman2022}, but they do not leverage local memory in order to bypass the cache coherence issue.

\section{Buffer Management}\label{sec:buf}
Accessing local memory in a compute node is still faster than accessing RDMA-enabled remote memory in DSM. Thus, it makes sense to cache hot data in the limited local memory (of compute nodes) to minimize remote memory accesses. But there are unique challenges for the buffer management in \sys{}.

\textbf{Challenge \#8: Designing Light-weight Buffer Management}. 
Existing buffer management is optimized for the hierarchy where there exists a huge performance gap between a cache hit and miss, e.g., the latency gap between main-memory and disk can be 100,000$\times$. Thus, the goal of existing buffer management is to improve the cache hit rates by developing optimized techniques, e.g., optimizing buffer replacement policies~\cite{LRUK93,LIRS02,MLCache19,ARC03} and storing compressed pages in the buffer~\cite{SQLServerCompress2009}. In \sys{}, we need to rethink  buffer management because the performance gap between local and remote memory is significantly narrowed, e.g., down to 10$\times$ or less, due to fast RDMA networking. Thus, we need to focus on the actual running time instead of just cache hit rates. That is because, software overhead, e.g., lookup cost, maintenance cost to reorganize buffer contents (in, say LRU), and synchronization cost due to multi-threaded access may become the performance bottlenecks for fast RDMA. These have traditionally not been major concerns for slower devices, e.g., SSDs or HDDs.
Thus, research is needed to evaluate the overhead of popular buffer management policies, e.g., LRU, LRU-K~\cite{LRUK93}, 2Q~\cite{TwoQ94}, CLOCK, and ARC~\cite{ARC03}. New buffer management policies must consider actual running time instead of purely optimizing cache hit rates.

Buffer management in \sys{} is different from the buffer management optimized for the hierarchy of local memory and local persistent memory (PM)~\cite{PMbuffer18,Spitfire21}, because CPU can directly operate on data stored on the local PM if there is a cache miss. However, in \sys{}, data must be transferred from remote memory to local memory first before being accessed if a cache miss happens.

Another direction is to evaluate the effectiveness of caching compressed pages. Depending on different data types and compression techniques,  decompression overhead might even be higher than directly fetching uncompressed data from remote memory. Thus, light-weight compression is important for \sys{}.

\textbf{Challenge \#9: Caching vs. Offloading}. Caching and offloading are two popular techniques to improve the performance in disaggregated databases. A recent study~\cite{FlexPushdownDB21} shows that caching and offloading are not orthogonal to each other on storage-disaggregated databases with conventional TCP/IP networking~\cite{FlexPushdownDB21}. It is important to re-investigate their interaction within \sys{} for the following reasons: (1)~\sys{} uses RDMA that may make caching more favorable. Intuitively, if network latency is zero, it is favorable to bring data from remote memory to local memory upon a cache miss because compute nodes have better compute power. (2)~The analytical model in \cite{FlexPushdownDB21} has high software overhead to decide when to cache or offload, while \sys{} requires a light-weight model. (3)~\sys{} targets OLTP applications that involve updates, which can incur inconsistencies between the cached data and the underlying data (due to pushing down).

{\bf Existing Research}. 
Existing work on MD has not focused  on the challenges of buffer management mentioned above because they still use disk-based buffer management for MD, e.g., \cite{ZhangCIDR20,ZhangVLDB20,PolarDBServerless21,Disaggregation21}. There are other works on using remote memory as a buffer of disk, e.g., \cite{CompuCache22,LiDSN16}. However, this work focuses on the buffer management for the two-level hierarchy with local memory and remote memory (without involving disk) where all the data is stored in remote memory with hot data being cached in local memory.

\vspace{-10pt}
\section{Index Design}\label{sec:index}
RDMA-based MD also has profound implications on index design for \sys{} that we highlight in this section.

\begin{sloppypar}

\textbf{Challenge \#10: RDMA-Conscious Index Design}. Index design needs to be hardware conscious to truly achieve good performance. All state of the art indexes used in modern systems heavily rely on hardware conscious design, e.g.,  Bw-tree~\cite{Bwtree13}, Masstree~\cite{Masstree12}, ART~\cite{ART13}, and LSM-tree~\cite{LSMTree96,DayanAI17}. In \sys{}, compute nodes access remote memory, i.e., the DSM layer, via RDMA. The intrinsic properties of RDMA networking need to be at the core of index design. There are numerous new hardware related design factors to explore that do not always have an equivalent context with past hardware properties. These factors include: (1)~Which RDMA primitive to use, e.g., one- or two-sided RDMA, synchronous or asynchronous RDMA; (2)~How to best utilize the limited buffer memory in each compute node; (3)~How to reduce software overhead to best leverage high-performance RDMA; and (4)~How to exploit the computing capability of memory nodes to reduce data transfer.
\end{sloppypar}

These design choices are not independent, e.g., having bigger local memory gives preference to using  one-sided RDMA in favor of fewer round trips. Also, using near-data computing impacts the way of using buffers.
Another direction is to design indexes that adaptively balance available compute, memory, and RDMA resources to prevent imbalanced resource utilization, e.g., a compute node's CPU is fully saturated while a memory node's CPU or RDMA bandwidth are largely idle.

While we may able to utilize and build on past research in many cases, in other cases it might be necessary to reconsider and drop design choices that are considered state of the art. For example, if we can access remote nodes extremely fast, then approaches that rely on sketches and summaries to filter data remotely may prove less impactful in this context. 

\textbf{Challenge \#11: Concurrent Index Operations}. 
How to handle concurrent accesses (reads/writes) from a very large number of compute nodes? We expect a stronger requirement for concurrency with MD because more nodes (and thus more queries) will have access to the same memory on shared data copies. Although existing indexes support multi-threaded access via lock-based or lock-free designs, e.g., Bw-tree~\cite{Bwtree13}, it is unclear how they perform in \sys{} due to (1)~Expensive RDMA locking overhead; and (2)~No hardware-supported cache coherence. Also, for lock-based approaches, e.g., ART~\cite{ART13}, it is unclear what lock types to use. 
Thus, we need to design new multi-threaded indexes that optimize for multiple compute nodes accessing DSM via RDMA. 

Besides that, LSM-based indexing~\cite{LSMTree96,DayanAI17} can be worth investigating because 
it naturally fits the local memory and remote memory hierarchy. For example, LSM-trees can hold filters and fence pointers in compute nodes as they help protect from unnecessary round trips. 
More research is needed to reduce software overhead and leverage the compute capabilities of memory nodes, e.g., offloading LSM compaction to memory nodes.

{\bf Existing Research}. Sherman~\cite{Sherman2022} is an optimized B-tree for MD. It uses one-sided RDMA to access remote memory, and addresses  concurrent accesses using RDMA-based exclusive locks and version validation. To reduce network round trips (due to one-sided RDMA), Sherman~\cite{Sherman2022} caches all  internal nodes into local memory, which  consumes more memory. Moreover, Sherman does not leverage the compute resources in memory nodes.
Ziegler et al. propose an RDMA-optimized B-tree structure that spans multiple memory nodes~\cite{RDMABtree19}, but it does not target MD and does not use the compute nodes' local memory. 
RACE is a  hash index for MD~\cite{ZuoSYZ021} but it only uses one-sided RDMA. It implements a lock-free multi-node CC protocol for the hash buckets. Overall, while there are a few prior works targeting MD, there is still big room for improvement because existing works have not fully leveraged RDMA characteristics. 

\section{Discussion}\label{sec:drawbacks}

\textbf{Performance}. 
The performance of \sys{} due to MD can be mitigated by several approaches. 
(1)~Use more local memory in each compute node. As demonstrated in~\cite{Disaggregation21}, caching 50\% data in local memory achieves almost no performance drop. Obviously, there is a tradeoff between more local memory capacity and memory utilization. But MD introduces  flexibility in controlling local memory size and can break the memory capacity limit of a single server. (2)~Optimize buffer management as in \sect{}~\ref{sec:buf}. (3)~Send only logs (i.e., log-as-the-database) as in Aurora~\cite{Aurora17}. (4)~Leverage near-data computing in the shared-memory layer to reduce data movement~\cite{Qizhen22}.


\begin{sloppypar}
\textbf{Distributed Shared-Nothing vs. DSM}. 
For the main-memory DBMSs considered in this paper, there are the DSN-DBs and DSM-DBs choices, so what is the difference between the two, really?

It is true that DSN-DBs can benefit from fast RDMA to reduce network communication cost, but maybe not too much because DSN-DBs are purposely designed for slow networks by carefully developing techniques to localize transactions and query processing to minimize network data accesses. With RDMA, one can relax the network accesses by allowing every node in DSN-DBs to access the main-memory of every node, which somehow mimics DSM accesses. However, that is not the \sys{} this paper is advocating for because that does not support memory elasticity and independent memory scaling. In order to do so, we need to have asymmetric architecture of "compute nodes" and "memory nodes" with different compute and memory capabilities. Also, the memory nodes form a DSM layer, which becomes \sys{}'s proposal. A benchmark that systematically compares the DSN-DBs and DSM-DBs is required to best understand the two designs. We believe that DSM-DBs can better leverage fast RDMA networking than DSN-DBs. 
\end{sloppypar}

Observe that DSN-DBs and DSM-DBs are not incompatible. For large-scale applications that require cross data-center deployment, DSM-DBs alone would not work because RDMA is not applicable due to the long latency dominated by speed-of-light delays among data-centers. Thus, a hybrid design that combines shared-memory and shared-nothing is required with shared-memory within the same data center and shared-nothing across data centers.

Note that we skipped the discussion of distributed shared-storage (DSS) since this paper focuses on main-memory databases. But if we involve storage, then DSS-DBs are implied in DSM-DBs. The above discussions hold between DSM-DBs and DSN-DBs with storage.

\section{Related Work}\label{sec:related}

\textbf{Distributed Shared-Storage Databases (DSS-DB)}. \sys{} has the potential to address several issues that are found challenging in DSS-DBs, e.g., Aurora~\cite{Aurora17}, Socrates~\cite{Socrates19}, PolarDB~\cite{PolarFS18}, and Taurus~\cite{Taurus20}. (1) \sys{} can address the multi-master issue -- a challenging issue in DSS-DBs -- by leveraging fast RDMA and the DSM layer to quickly synchronize the states between compute nodes. Note that although PolarDB uses RDMA~\cite{PolarFS18}, it does not efficiently support multi-master because a compute node cannot use one-sided RDMA to access the storage nodes. (2) \sys{} can easily support memory elasticity, independent scaling and failure of compute and memory that are not possible in DSS-DBs.

\textbf{Distributed Shared-Nothing Databases (DSN-DB)}. In addition to supporting memory elasticity and independent scaling, \sys{} has the potential to efficiently address the distributed transaction issues in DSN-DBs (e.g., VoltDB~\cite{VoltDB}, MemSQL~\cite{MemSQL16}, and Hekaton~\cite{Hekaton13}), because the RDMA-connected DSM layer in \sys{} provides a fast way to reshard data among compute nodes. This makes \sys{} more resilient to skew due to fast resharding.

\begin{sloppypar}
\textbf{Distributed Main-Memory Databases}. \sys{} is different from existing distributed main-memory databases (e.g., VoltDB~\cite{VoltDB}, MemSQL~\cite{MemSQL16}, and Hekaton~\cite{Hekaton13}) because \sys{} is shared-memory while these databases are shared-nothing. Simply adding RDMA networking to these shared-nothing main-memory databases does not solve the memory elasticity and independent scaling problems. It is required to have separated notions of "compute nodes" and "memory nodes" and allow compute nodes to access memory nodes via abstract APIs, which is \sys{}'s proposal.
\end{sloppypar}

\textbf{NAM (Network-Attached Memory)~\cite{ZamanianBKH17,BinnigCGKZ16}}. NAM is an innovative architecture that allows compute nodes to access a shared-memory pool of memory nodes. However, NAM is not designed for MD. The compute and memory nodes in NAM are \textit{logically decoupled}, while \sys{} emphasizes \textit{physical decoupling} due to MD. Although logical decoupling has the potential to co-locate compute and memory nodes (where each "node" is a process) to reduce network access, physical decoupling enables additional benefits, e.g., independent failure and crash handling of compute and memory nodes, and better resource utilization and elasticity. Besides that, this vision paper has a wider scope that also includes durability, availability, DSM APIs, buffer management, and index design.

\textbf{Impact of MD on Databases}. Recent research investigates the impact of MD on databases, e.g.,~\cite{ZhangCIDR20,ZhangVLDB20,Farview22,Qizhen22}. These works target single-node databases (with a single compute node and a single memory node) instead of a distributed DBMS as in \sys{}. Also, they focus on OLAP databases while \sys{} focuses on OLTP databases. PolarDB incorporates MD~\cite{PolarDBServerless21,Disaggregation21}, but is still a disk-based DBMS with a single master node. In contrast, \sys{} is main-memory-based that supports multi-masters, where every compute node can process read/write requests to improve scalability.

\section{Conclusion}\label{sec:conclusion}
Memory disaggregation (MD) is regarded as the next technology breakthrough by major tech companies. This paper presents our vision on the impact of MD to distributed databases, in particular OLTP main-memory databases. We envision that the distributed shared-memory (DSM) architecture that has been under-appreciated in the past can be promising in the future due to MD. This paper highlights new problems and challenges in realizing \sys{} with memory disaggregation over RDMA. 


\clearpage\newpage
\balance
\bibliographystyle{abbrv} 
\bibliography{paper}

\begin{thebibliography}{10}

\bibitem{IBMCloud}
{Advancing Cloud with Memory Disaggregation,
  \url{https://www.ibm.com/blogs/research/2018/01/advancing-cloud-memory-disaggregation/}}.

\bibitem{AlloyDB}
{AlloyDB for PostgreSQL, \url{https://cloud.google.com/alloydb}}.

\bibitem{AWSEBS}
{Amazon EBS, \url{https://aws.amazon.com/ebs/features/}}.

\bibitem{AWSS3}
{Amazon S3, \url{https://aws.amazon.com/pm/serv-s3/}}.

\bibitem{IntelRSD}
{Intel RSD,
  \url{https://www.intel.com/content/www/us/en/architecture-and-technology/rack-scale-design-overview.html}}.

\bibitem{Mellanox}
{Mellanox Connectx-6,
  \url{https://www.mellanox.com/products/ethernet-adapter-ic/connectx-6-en-ic}}.

\bibitem{Memcached}
{Memcached, \url{https://memcached.org/}}.

\bibitem{VoltDB}
{VoltDB, \url{https://www.voltdb.com/}}.

\bibitem{AbebeGD20}
M.~Abebe, B.~Glasbergen, and K.~Daudjee.
\newblock {DynaMast: Adaptive Dynamic Mastering for Replicated Systems}.
\newblock In {\em International Conference on Data Engineering (ICDE)}, pages
  1381--1392, 2020.

\bibitem{POLARIS20}
J.~Aguilar{-}Saborit and R.~Ramakrishnan.
\newblock {POLARIS: The Distributed {SQL} Engine in Azure Synapse}.
\newblock {\em Proceedings of the VLDB Endowment (PVLDB)}, 13(12):3204--3216,
  2020.

\bibitem{Socrates19}
P.~Antonopoulos, A.~Budovski, C.~Diaconu, A.~H. Saenz, J.~Hu, H.~Kodavalla,
  D.~Kossmann, S.~Lingam, U.~F. Minhas, N.~Prakash, V.~Purohit, H.~Qu, C.~S.
  Ravella, K.~Reisteter, S.~Shrotri, D.~Tang, and V.~Wakade.
\newblock {Socrates: The New SQL Server in the Cloud}.
\newblock In {\em Proceedings of the ACM International Conference on Management
  of Data (SIGMOD)}, pages 1743--1756, 2019.

\bibitem{ArulrajPP16}
J.~Arulraj, M.~Perron, and A.~Pavlo.
\newblock {Write-Behind Logging}.
\newblock {\em Proceedings of the VLDB Endowment (PVLDB)}, 10(4):337--348,
  2016.

\bibitem{BangMPB20}
T.~Bang, N.~May, I.~Petrov, and C.~Binnig.
\newblock {The Tale of 1000 Cores: An Evaluation of Concurrency Control on
  Real(ly) Large Multi-socket Hardware}.
\newblock In {\em Proceedings of the International Workshop on Data Management
  on New Hardware (DaMoN)}, pages 3:1--3:9, 2020.

\bibitem{BinnigCGKZ16}
C.~Binnig, A.~Crotty, A.~Galakatos, T.~Kraska, and E.~Zamanian.
\newblock {The End of Slow Networks: It's Time for a Redesign}.
\newblock {\em Proceedings of the VLDB Endowment (PVLDB)}, 9(7):528--539, 2016.

\bibitem{GAM18}
Q.~Cai, W.~Guo, H.~Zhang, D.~Agrawal, G.~Chen, B.~C. Ooi, K.~Tan, Y.~M. Teo,
  and S.~Wang.
\newblock {Efficient Distributed Memory Management with RDMA and Caching}.
\newblock {\em Proceedings of the VLDB Endowment (PVLDB)}, 11(11):1604--1617,
  2018.

\bibitem{PolarFS18}
W.~Cao, Z.~Liu, P.~Wang, S.~Chen, C.~Zhu, S.~Zheng, Y.~Wang, and G.~Ma.
\newblock {PolarFS: An Ultra-Low Latency and Failure Resilient Distributed File
  System for Shared Storage Cloud Database}.
\newblock {\em Proceedings of the VLDB Endowment (PVLDB)}, 11(12):1849--1862,
  2018.

\bibitem{PolarDBServerless21}
W.~Cao, Y.~Zhang, X.~Yang, F.~Li, S.~Wang, Q.~Hu, X.~Cheng, Z.~Chen, Z.~Liu,
  J.~Fang, B.~Wang, Y.~Wang, H.~Sun, Z.~Yang, Z.~Cheng, S.~Chen, J.~Wu, W.~Hu,
  J.~Zhao, Y.~Gao, S.~Cai, Y.~Zhang, and J.~Tong.
\newblock {PolarDB Serverless: A Cloud Native Database for Disaggregated Data
  Centers}.
\newblock In {\em Proceedings of the ACM International Conference on Management
  of Data (SIGMOD)}, pages 2477--2489, 2021.

\bibitem{MemSQL16}
J.~Chen, S.~Jindel, R.~Walzer, R.~Sen, N.~Jimsheleishvilli, and M.~Andrews.
\newblock {The MemSQL Query Optimizer: {A} Modern Optimizer for Real-time
  Analytics in a Distributed Database}.
\newblock {\em Proceedings of the VLDB Endowment (PVLDB)}, 9(13):1401--1412,
  2016.

\bibitem{Granola12}
J.~A. Cowling and B.~Liskov.
\newblock {Granola: Low-Overhead Distributed Transaction Coordination}.
\newblock In {\em USENIX Annual Technical Conference (ATC)}, pages 223--235,
  2012.

\bibitem{Citus21}
U.~Cubukcu, O.~Erdogan, S.~Pathak, S.~Sannakkayala, and M.~Slot.
\newblock {Citus: Distributed PostgreSQL for Data-Intensive Applications}.
\newblock In {\em Proceedings of the ACM International Conference on Management
  of Data (SIGMOD)}, pages 2490--2502, 2021.

\bibitem{Snowflake16}
B.~Dageville, T.~Cruanes, M.~Zukowski, V.~Antonov, A.~Avanes, J.~Bock,
  J.~Claybaugh, D.~Engovatov, M.~Hentschel, J.~Huang, A.~W. Lee, A.~Motivala,
  A.~Q. Munir, S.~Pelley, P.~Povinec, G.~Rahn, S.~Triantafyllis, and
  P.~Unterbrunner.
\newblock {The Snowflake Elastic Data Warehouse}.
\newblock In {\em Proceedings of the ACM International Conference on Management
  of Data (SIGMOD)}, pages 215--226, 2016.

\bibitem{DayanAI17}
N.~Dayan, M.~Athanassoulis, and S.~Idreos.
\newblock {Monkey: Optimal Navigable Key-Value Store}.
\newblock In {\em Proceedings of the ACM International Conference on Management
  of Data (SIGMOD)}, pages 79--94, 2017.

\bibitem{Taurus20}
A.~Depoutovitch, C.~Chen, J.~Chen, P.~Larson, S.~Lin, J.~Ng, W.~Cui, Q.~Liu,
  W.~Huang, Y.~Xiao, and Y.~He.
\newblock {Taurus Database: How to be Fast, Available, and Frugal in the
  Cloud}.
\newblock In {\em Proceedings of the ACM International Conference on Management
  of Data (SIGMOD)}, pages 1463--1478, 2020.

\bibitem{DeWittKOSSW84}
D.~J. DeWitt, R.~H. Katz, F.~Olken, L.~D. Shapiro, M.~Stonebraker, and D.~A.
  Wood.
\newblock {Implementation Techniques for Main Memory Database Systems}.
\newblock In {\em Proceedings of the ACM International Conference on Management
  of Data (SIGMOD)}, pages 1--8, 1984.

\bibitem{Hekaton13}
C.~Diaconu, C.~Freedman, E.~Ismert, P.~Larson, P.~Mittal, R.~Stonecipher,
  N.~Verma, and M.~Zwilling.
\newblock {Hekaton: SQL Server's Memory-Optimized OLTP Engine}.
\newblock In {\em Proceedings of the ACM International Conference on Management
  of Data (SIGMOD)}, pages 1243--1254, 2013.

\bibitem{FaRM14}
A.~Dragojevic, D.~Narayanan, M.~Castro, and O.~Hodson.
\newblock {FaRM: Fast Remote Memory}.
\newblock In {\em Proceedings of the USENIX Symposium on Networked Systems
  Design and Implementation (NSDI)}, pages 401--414, 2014.

\bibitem{MMDB17}
F.~Faerber, A.~Kemper, P.~Larson, J.~J. Levandoski, T.~Neumann, and A.~Pavlo.
\newblock {Main Memory Database Systems}.
\newblock {\em Foundations and Trends in Databases}, 8(1-2):1--130, 2017.

\bibitem{GarciaMolinaS92}
H.~Garcia{-}Molina and K.~Salem.
\newblock {Main Memory Database Systems: An Overview}.
\newblock {\em IEEE Transactions on Knowledge and Data Engineering (TKDE)},
  4(6):509--516, 1992.

\bibitem{EvaDT17}
R.~Harding, D.~V. Aken, A.~Pavlo, and M.~Stonebraker.
\newblock {An Evaluation of Distributed Concurrency Control}.
\newblock {\em Proceedings of the VLDB Endowment (PVLDB)}, 10(5):553--564,
  2017.

\bibitem{LIRS02}
S.~Jiang and X.~Zhang.
\newblock {LIRS: An Efficient Low Inter-Reference Recency Set Replacement
  Policy to Improve Buffer Cache Performance}.
\newblock In {\em Proceedings of the International Conference on Measurement
  and Modeling of Computer Systems (SIGMETRICS)}, pages 31--42, 2002.

\bibitem{TwoQ94}
T.~Johnson and D.~E. Shasha.
\newblock {2Q: {A} Low Overhead High Performance Buffer Management Replacement
  Algorithm}.
\newblock In {\em International Conference on Very Large Data Bases (VLDB)},
  pages 439--450, 1994.

\bibitem{HPTheMachine}
K.~Keeton.
\newblock {Memory-Driven Computing.
  \url{https://www.usenix.org/sites/default/files/conference/protected-files/fast17_slides_keeton.pdf}}.
\newblock In {\em USENIX Conference on File and Storage Technologies (FAST)},
  2017.

\bibitem{Farview22}
D.~Korolija, D.~Koutsoukos, K.~Keeton, K.~Taranov, D.~S. Milojicic, and
  G.~Alonso.
\newblock {Farview: Disaggregated Memory with Operator Off-loading for Database
  Engines}.
\newblock In {\em Conference on Innovative Data Systems Research (CIDR)}, 2022.

\bibitem{Hydra22}
Y.~Lee, H.~A. Maruf, M.~Chowdhury, A.~Cidon, and K.~G. Shin.
\newblock {Hydra: Resilient and Highly Available Remote Memory}.
\newblock In {\em USENIX Conference on File and Storage Technologies (FAST)},
  pages 181--197, 2022.

\bibitem{ART13}
V.~Leis, A.~Kemper, and T.~Neumann.
\newblock {The Adaptive Radix Tree: ARTful Indexing for Main-memory Databases}.
\newblock In {\em International Conference on Data Engineering (ICDE)}, pages
  38--49, 2013.

\bibitem{Bwtree13}
J.~J. Levandoski, D.~B. Lomet, and S.~Sengupta.
\newblock {The Bw-Tree: A B-tree for New Hardware Platforms}.
\newblock In {\em International Conference on Data Engineering (ICDE)}, pages
  302--313, 2013.

\bibitem{LiDSN16}
F.~Li, S.~Das, M.~Syamala, and V.~R. Narasayya.
\newblock {Accelerating Relational Databases by Leveraging Remote Memory and
  RDMA}.
\newblock In {\em Proceedings of the ACM International Conference on Management
  of Data (SIGMOD)}, pages 355--370, 2016.

\bibitem{AzureMD2022}
H.~Li, D.~S. Berger, S.~Novakovic, L.~Hsu, D.~Ernst, P.~Zardoshti, M.~Shah,
  I.~Agarwal, M.~D. Hill, M.~Fontoura, and R.~Bianchini.
\newblock {First-generation Memory Disaggregation for Cloud Platforms}.
\newblock {\em CoRR}, abs/2203.00241, 2022.

\bibitem{Kai89}
K.~Li and P.~Hudak.
\newblock {Memory Coherence in Shared Virtual Memory Systems}.
\newblock {\em ACM Transactions on Computer Systems (TOCS)}, 7(4):321--359,
  1989.

\bibitem{LinC0OTW16}
Q.~Lin, P.~Chang, G.~Chen, B.~C. Ooi, K.~Tan, and Z.~Wang.
\newblock {Towards a Non-2PC Transaction Management in Distributed Database
  Systems}.
\newblock In {\em Proceedings of the ACM International Conference on Management
  of Data (SIGMOD)}, pages 1659--1674, 2016.

\bibitem{MalviyaWMS14}
N.~Malviya, A.~Weisberg, S.~Madden, and M.~Stonebraker.
\newblock {Rethinking Main Memory {OLTP}} recovery.
\newblock In {\em International Conference on Data Engineering (ICDE)}, pages
  604--615, 2014.

\bibitem{Masstree12}
Y.~Mao, E.~Kohler, and R.~T. Morris.
\newblock {Cache Craftiness for Fast Multicore Key-value Storage}.
\newblock In {\em European Conference on Computer Systems (EuroSys)}, pages
  183--196, 2012.

\bibitem{ARC03}
N.~Megiddo and D.~S. Modha.
\newblock {ARC: {A} Self-Tuning, Low Overhead Replacement Cache}.
\newblock In {\em USENIX Conference on File and Storage Technologies (FAST)},
  pages 115 -- 130, 2003.

\bibitem{SQLServerCompress2009}
S.~Mishra.
\newblock {Data Compression: Strategy, Capacity Planning and Best Practices,
  \url{https://docs.microsoft.com/en-us/previous-versions/sql/sql-server-2008/dd894051(v=sql.100)}},
  2009.

\bibitem{NitzbergL91}
B.~Nitzberg and V.~M. Lo.
\newblock {Distributed Shared Memory: {A} Survey of Issues and Algorithms}.
\newblock {\em Computer}, 24(8):52--60, 1991.

\bibitem{LRUK93}
E.~J. O'Neil, P.~E. O'Neil, and G.~Weikum.
\newblock {The {LRU-K} Page Replacement Algorithm For Database Disk Buffering}.
\newblock In {\em Proceedings of the ACM International Conference on Management
  of Data (SIGMOD)}, pages 297--306, 1993.

\bibitem{LSMTree96}
P.~E. O'Neil, E.~Cheng, D.~Gawlick, and E.~J. O'Neil.
\newblock {The Log-Structured Merge-Tree (LSM-Tree)}.
\newblock {\em Acta Informatica}, 33(4):351--385, 1996.

\bibitem{OngaroRSOR11}
D.~Ongaro, S.~M. Rumble, R.~Stutsman, J.~K. Ousterhout, and M.~Rosenblum.
\newblock {Fast Crash Recovery in RAMCloud}.
\newblock In {\em Proceedings of the Symposium on Operating Systems Principles
  (SOSP)}, pages 29--41, 2011.

\bibitem{ramcloud09}
J.~K. Ousterhout, P.~Agrawal, D.~Erickson, C.~Kozyrakis, J.~Leverich,
  D.~Mazi{\`{e}}res, S.~Mitra, A.~Narayanan, G.~M. Parulkar, M.~Rosenblum,
  S.~M. Rumble, E.~Stratmann, and R.~Stutsman.
\newblock {The Case for RAMClouds: Scalable High-performance Storage Entirely
  in {DRAM}}.
\newblock {\em ACM SIGOPS Operating Systems Review}, 43(4):92--105, 2009.

\bibitem{OzsuV14}
M.~T. {\"{O}}zsu and P.~Valduriez.
\newblock {\em {Distributed and Parallel Database Systems, Third Edition}}.
\newblock {CRC} Press, 2014.

\bibitem{ProticTM96}
J.~Protic, M.~Tomasevic, and V.~Milutinovic.
\newblock {Distributed Shared Memory: Concepts and Systems}.
\newblock {\em IEEE Parallel \& Distributed Technology: Systems \&
  Applications}, 4(2):63--71, 1996.

\bibitem{SathiamoorthyAPDVCB13}
M.~Sathiamoorthy, M.~Asteris, D.~S. Papailiopoulos, A.~G. Dimakis, R.~Vadali,
  S.~Chen, and D.~Borthakur.
\newblock {XORing Elephants: Novel Erasure Codes for Big Data}.
\newblock {\em Proceedings of the VLDB Endowment (PVLDB)}, 6(5):325--336, 2013.

\bibitem{MLCache19}
Z.~Shi, X.~Huang, A.~Jain, and C.~Lin.
\newblock {Applying Deep Learning to the Cache Replacement Problem}.
\newblock In {\em Proceedings of the International Symposium on
  Microarchitecture (MICRO)}, pages 413--425, 2019.

\bibitem{StonebrakerSharedNothing86}
M.~Stonebraker.
\newblock {The Case for Shared Nothing}.
\newblock {\em IEEE Data Engineering Bulletin}, 9(1):4--9, 1986.

\bibitem{StonebrakerSharedNothing2011}
M.~Stonebraker.
\newblock {Shared-nothing vs Shared-disk,
  \url{https://www.youtube.com/watch?v=G-o2bFd91Sw}}.
\newblock In {\em Extremely Large Databases Workshop (XLDB)}, 2011.

\bibitem{TanabeHKT20}
T.~Tanabe, T.~Hoshino, H.~Kawashima, and O.~Tatebe.
\newblock {An Analysis of Concurrency Control Protocols for In-Memory Database
  with CCBench}.
\newblock {\em Proceedings of the VLDB Endowment (PVLDB)}, 13(13):3531--3544,
  2020.

\bibitem{TaranovGH21}
K.~Taranov, S.~D. Girolamo, and T.~Hoefler.
\newblock {CoRM: Compactable Remote Memory over {RDMA}}.
\newblock In {\em Proceedings of the ACM International Conference on Management
  of Data (SIGMOD)}, pages 1811--1824, 2021.

\bibitem{Calvin12}
A.~Thomson, T.~Diamond, S.~Weng, K.~Ren, P.~Shao, and D.~J. Abadi.
\newblock {Calvin: Fast Distributed Transactions for Partitioned Database
  Systems}.
\newblock In {\em Proceedings of the ACM International Conference on Management
  of Data (SIGMOD)}, pages 1--12, 2012.

\bibitem{PMbuffer18}
A.~van Renen, V.~Leis, A.~Kemper, T.~Neumann, T.~Hashida, K.~Oe, Y.~Doi,
  L.~Harada, and M.~Sato.
\newblock {Managing Non-Volatile Memory in Database Systems}.
\newblock In {\em Proceedings of the ACM International Conference on Management
  of Data (SIGMOD)}, pages 1541--1555, 2018.

\bibitem{Aurora17}
A.~Verbitski, A.~Gupta, D.~Saha, M.~Brahmadesam, K.~Gupta, R.~Mittal,
  S.~Krishnamurthy, S.~Maurice, T.~Kharatishvili, and X.~Bao.
\newblock {Amazon Aurora: Design Considerations for High Throughput
  Cloud-Native Relational Databases}.
\newblock In {\em Proceedings of the ACM International Conference on Management
  of Data (SIGMOD)}, pages 1041--1052, 2017.

\bibitem{wang2021rdma}
C.~Wang and X.~Qian.
\newblock {RDMA-enabled Concurrency Control Protocols for Transactions in the
  Cloud Era}.
\newblock {\em IEEE Transactions on Cloud Computing}, 2021.

\bibitem{Sherman2022}
Q.~Wang, Y.~Lu, and J.~Shu.
\newblock {Sherman: {A} Write-Optimized Distributed B+Tree Index on
  Disaggregated Memory}.
\newblock In {\em Proceedings of the ACM International Conference on Management
  of Data (SIGMOD)}, pages 1033--1048, 2022.

\bibitem{MOCC16}
T.~Wang and H.~Kimura.
\newblock {Mostly-Optimistic Concurrency Control for Highly Contended Dynamic
  Workloads on a Thousand Cores}.
\newblock {\em Proceedings of the VLDB Endowment (PVLDB)}, 10(2):49--60, 2016.

\bibitem{FlexPushdownDB21}
Y.~Yang, M.~Youill, M.~E. Woicik, Y.~Liu, X.~Yu, M.~Serafini, A.~Aboulnaga, and
  M.~Stonebraker.
\newblock {FlexPushdownDB: Hybrid Pushdown and Caching in a Cloud {DBMS}}.
\newblock {\em Proceedings of the VLDB Endowment (PVLDB)}, 14(11):2101--2113,
  2021.

\bibitem{AbyssCC2014}
X.~Yu, G.~Bezerra, A.~Pavlo, S.~Devadas, and M.~Stonebraker.
\newblock {Staring into the Abyss: An Evaluation of Concurrency Control with
  One Thousand Cores}.
\newblock {\em Proceedings of the VLDB Endowment (PVLDB)}, 8(3):209--220, 2014.

\bibitem{ZamanianBKH17}
E.~Zamanian, C.~Binnig, T.~Kraska, and T.~Harris.
\newblock {The End of a Myth: Distributed Transaction Can Scale}.
\newblock {\em Proceedings of the VLDB Endowment (PVLDB)}, 10(6):685--696,
  2017.

\bibitem{AnalyticDB19}
C.~Zhan, M.~Su, C.~Wei, X.~Peng, L.~Lin, S.~Wang, Z.~Chen, F.~Li, Y.~Pan,
  F.~Zheng, and C.~Chai.
\newblock {AnalyticDB: Real-time OLAP Database System at Alibaba Cloud}.
\newblock {\em Proceedings of the VLDB Endowment (PVLDB)}, 12(12):2059--2070,
  2019.

\bibitem{Redy22}
Q.~Zhang, P.~A. Bernstein, D.~S. Berger, and B.~Chandramouli.
\newblock {Redy: Remote Dynamic Memory Cache}.
\newblock {\em Proceedings of the VLDB Endowment (PVLDB)}, 15(4):766 -- 779,
  2022.

\bibitem{CompuCache22}
Q.~Zhang, P.~A. Bernstein, D.~S. Berger, B.~Chandramouli, V.~Liu, and B.~T.
  Loo.
\newblock {CompuCache: Remote Computable Caching using Spot VMs}.
\newblock In {\em Conference on Innovative Data Systems Research (CIDR)}, 2022.

\bibitem{ZhangCIDR20}
Q.~Zhang, Y.~Cai, S.~Angel, V.~Liu, A.~Chen, and B.~T. Loo.
\newblock {Rethinking Data Management Systems for Disaggregated Data Centers}.
\newblock In {\em Conference on Innovative Data Systems Research (CIDR)}, 2020.

\bibitem{ZhangVLDB20}
Q.~Zhang, Y.~Cai, X.~Chen, S.~Angel, A.~Chen, V.~Liu, and B.~T. Loo.
\newblock {Understanding the Effect of Data Center Resource Disaggregation on
  Production DBMSs}.
\newblock {\em Proceedings of the VLDB Endowment (PVLDB)}, 13(9):1568--1581,
  2020.

\bibitem{Qizhen22}
Q.~Zhang, X.~Chen, S.~Sankhe, Z.~Zheng, K.~Zhong, S.~Angel, A.~Chen, V.~Liu,
  and B.~T. Loo.
\newblock {Optimizing Data-intensive Systems in Disaggregated Data Centers with
  {TELEPORT}}.
\newblock In {\em Proceedings of the ACM International Conference on Management
  of Data (SIGMOD)}, pages 1345--1359, 2022.

\bibitem{Disaggregation21}
Y.~Zhang, C.~Ruan, C.~Li, J.~Yang, W.~Cao, F.~Li, B.~Wang, J.~Fang, Y.~Wang,
  J.~Huo, and C.~Bi.
\newblock {Towards Cost-Effective and Elastic Cloud Database Deployment via
  Memory Disaggregation}.
\newblock {\em Proceedings of the VLDB Endowment (PVLDB)}, 14(10):1900--1912,
  2021.

\bibitem{Spitfire21}
X.~Zhou, J.~Arulraj, A.~Pavlo, and D.~Cohen.
\newblock {Spitfire: {A} Three-Tier Buffer Manager for Volatile and
  Non-Volatile Memory}.
\newblock In {\em Proceedings of the ACM International Conference on Management
  of Data (SIGMOD)}, pages 2195--2207, 2021.

\bibitem{RDMABtree19}
T.~Ziegler, S.~T. Vani, C.~Binnig, R.~Fonseca, and T.~Kraska.
\newblock {Designing Distributed Tree-based Index Structures for Fast
  RDMA-capable Networks}.
\newblock In {\em Proceedings of the ACM International Conference on Management
  of Data (SIGMOD)}, pages 741--758, 2019.

\bibitem{ZuoSYZ021}
P.~Zuo, J.~Sun, L.~Yang, S.~Zhang, and Y.~Hua.
\newblock {One-sided RDMA-Conscious Extendible Hashing for Disaggregated
  Memory}.
\newblock In {\em USENIX Annual Technical Conference (ATC)}, pages 15--29,
  2021.

\end{thebibliography}

\end{document}